\documentclass[aps,prd,twocolumn,nofootinbib,showpacs]{revtex4}

\usepackage{graphicx,color}
\usepackage[colorlinks=true, linkcolor=blue, citecolor=blue, urlcolor=blue]{hyperref}

\begin{document}

\title{ A precise determination of the top-quark pole mass}

\author{Sheng-Quan Wang$^{1}$}
\email[email:]{sqwang@cqu.edu.cn}

\author{Xing-Gang Wu$^2$}
\email[email:]{wuxg@cqu.edu.cn}

\author{Zong-Guo Si$^3$}
\email[email:]{zgsi@sdu.edu.cn}

\author{Stanley J. Brodsky$^4$}
\email[email:]{sjbth@slac.stanford.edu}

\address{$^1$School of Science, Guizhou Minzu University, Guiyang 550025, P.R. China}
\address{$^2$Department of Physics, Chongqing University, Chongqing 401331, P.R. China}
\address{$^3$Department of Physics, Shandong University, Jinan, Shandong 250100, P.R. China}
\address{$^4$SLAC National Accelerator Laboratory, Stanford University, Stanford, California 94039, USA}

\date{\today}

\begin{abstract}

The Principle of Maximum Conformality (PMC) provides a systematic way to eliminate the renormalization scheme and renormalization scale uncertainties for high-energy processes. We have observed that by applying PMC scale-setting, one obtains comprehensive and self-consistent pQCD predictions for the top-quark pair total cross-section and the top-quark pair forward-backward asymmetry in agreement with the measurements at the Tevatron and LHC. As a step forward, in the present paper, we determine the top-quark pole mass via a detailed comparison of the top-quark pair cross-section with the measurements at the Tevatron and LHC. The results for the top-quark pole mass are $m_t=174.6^{+3.1}_{-3.2}$ GeV for the Tevatron with $\sqrt{S}=1.96$ TeV, $m_t=173.7\pm1.5$ GeV and $174.2\pm1.7$ GeV for the LHC with $\sqrt{S} = 7$ TeV and $8$ TeV, respectively. Those predictions agree with the average, $173.34\pm0.76$ GeV, obtained from various collaborations via direct measurements. The consistency of the pQCD predictions using the PMC with all of the collider measurements at different energies provides an important verification of QCD.


\end{abstract}

\maketitle

\section{Introduction}

The top-quark is the heaviest particle of the Standard Model (SM), and its mass is one of the fundamental parameters within the SM. The large top-quark mass implies a strong top-quark Yukawa coupling to the Higgs boson, playing a special role in testing the electroweak symmetry breaking mechanism and for the search of new physics beyond the SM. The top quark decays before hadronization, and one can determine its mass by directly measuring its decay products~\cite{Dalitz:1992np}. Such measurements allow for the direct extraction of the top-quark mass ($m_t$), which however, relies heavily on the detailed reconstruction of the kinematics and  reconstruction efficiency~\cite{Chatrchyan:2013haa, Aaboud:2016000}. In 2014, a combination of measurements of the top-quark mass performed by the CDF and D0 experiments at the Tevatron collider and the ATLAS and CMS experiments at the Large Hadron Collider (LHC) gives~\cite{ATLAS:2014wva}, $m_t=173.34\pm0.76$ GeV. The direct measurements are based on analysis techniques which use top-pair events provided by Monte Carlo (MC) simulation for different assumed values of the top quark mass. Applying those techniques to data yields a mass quantity corresponding to the top quark mass scheme implemented in the MC, thus it is referred as the ``MC mass".

Another important approach for extracting the top-quark mass is done by using detailed comparisons of the pQCD predictions with the corresponding measurements; this method is indirect, but it provides complementary information on the top quark compared to direct measurements. Theoretical arguments suggest that the top-quark MC mass is within $\sim 1$ GeV of its pole mass~\cite{Buckley:2011ms}, and thus its use has a negligible effect on the determination of pole mass~\cite{Khachatryan:2016mqs, Fleming:2007qr} \footnote{The position of the pole in the quark propagator is defined as its pole mass, and the on-shell quark propagator has no infrared divergences in perturbation theory, it thus provides a perturbative definition of the quark mass~\cite{Tarrach:1980up, Kronfeld:1998di}.}. Thus in our present calculations, we shall only extract the top-quark pole mass and as usual directly take the determined top-quark MC mass by the experimental groups as the value of the top-quark pole mass. Recently, such indirect extractions of $m_t$ from the top-quark pair production channels by various experimental collaborations have been performed, giving the pole value, $m_t=173.8^{+1.7}_{-1.8}$ GeV from CMS ~\cite{Khachatryan:2016mqs}, $m_t=172.8^{+3.4}_{-3.2}$ GeV from D0~\cite{Abazov:2016ekt}, and $m_t=172.9^{+2.5}_{-2.6}$ GeV from ATLAS~\cite{Aad:2014kva}.

A key goal for the indirect determinations is to have a precise theoretical prediction for the top-quark pair production cross-section in order to provide maximal constraints on $m_t$. Practically, one can first set an arbitrary initial renormalization scale to do the pQCD calculation, whose value only need to be large enough to ensure the applicability of the pQCD theory. Under conventional scale-setting, the renormalization scale is fixed to its initial value, which is usually chosen as the typical momentum flow of the process or the one to eliminate large logs in the perturbative series. More explicitly, it is conventional to take the renormalization scale in those predictions as the top-quark mass $m_{t}$ to eliminate the large logarithmic terms such as $\ln(\mu_r/m_{t})$; one then varies the renormalization scale over an arbitrary range such as $[m_{t}/2, 2m_{t}]$ to ascertain the uncertainty. At sufficiently high order, a small renormalization scale-dependent prediction may be achieved for global quantities such as the total cross-section. However, such small renormalization scale dependence of the resulting prediction is due to cancelations among different orders; the renormalization scale uncertainty for each order is still uncertain and could be very large. In fact, when one applies conventional scale-setting, the renormalization scheme- and initial renormalization scale- dependence is introduced at any fixed order.

The Principle of Maximum Conformality (PMC)~\cite{Wu:2013ei, Brodsky:2011ta, Brodsky:2011ig, Mojaza:2012mf, Brodsky:2013vpa} provides a systematic way to eliminate renormalization scheme-and-scale ambiguities. It generalizes the BLM scale setting procedure~\cite{Brodsky:1982gc} to all orders. As in QED~\cite{GellMann:1954fq}, one shifts the argument of the running coupling at each order in the pQCD series to absorb all occurrences of the $\beta$-function. In addition, a convergent pQCD series without factorial renormalon divergence can be obtained. The PMC predictions are renormalization-scheme independent at each order in $\alpha_s$, since all of the scheme-dependent $\{\beta_i\}$-terms in the QCD perturbative series have been resummed into the running couplings. The PMC satisfies renormalization group invariance and satisfies all of the self-consistency conditions of the renormalization group~\cite{Brodsky:2012ms, Wu:2014iba}, and it reduces in the $N_C\to 0$ Abelian limit~\cite{Brodsky:1997jk} to the standard Gell-Mann-Low method~\cite{GellMann:1954fq}. A number of PMC applications are summarized in the review~\cite{Wu:2015rga}; in each case the PMC works successfully and leads to improved agreement with experiment.

By applying PMC scale-setting, we can achieve optimal renormalization scales of the process and thus obtain precise predictions for the top-quark pair production cross-section without conventional renormalization scale uncertainty~\cite{Brodsky:2012rj, Brodsky:2012sz, Brodsky:2012ik, Wang:2014sua}. Because of the uncalculated high-order terms, there is residual renormalization scale dependence for the PMC prediction. However such residual renormalization scale dependence is generally small either due to the perturbative nature of the PMC scales or due to the fast convergence of the conformal pQCD series; e.g. we have found that the residual renormalization for top-pair production is negligibly small at the NNLO level. The PMC predictions for the top-quark pair forward-backward asymmetry are also in agreement with the corresponding CDF and D0 measurements~\cite{Wang:2015lna}, since it correctly assigns different renormalization scales in the one- and two- gluon exchange amplitudes.

In subsequent sections, we will determine the top-quark pole mass from a detailed comparison of the top-quark pair production cross-section  predicted by applying the PMC with the measured values obtained by the Tevatron and LHC experiments. \\

\section{Top-quark pair production at the hadron colliders and the determination of top-quark pole mass}

The hadronic cross-section for the top-quark pair production can be written as the convolution of the factorized partonic cross-section $\hat{\sigma}_{ij}$ with the parton luminosities ${\cal L}_{ij}$:
\begin{eqnarray}
&&\sigma_{H_1 H_2 \to {t\bar{t} X}} \nonumber\\
&=& \sum_{i,j} \int\limits_{4m_{t}^2}^{S}\, ds \,\, {\cal L}_{ij}(s, S, \mu_f) \hat \sigma_{ij}(s,\alpha_s(\mu_r),\mu_r,\mu_f),
\end{eqnarray}
where
\begin{displaymath}
{\cal L}_{ij}(s, S, \mu_f) = {1\over S} \int\limits_s^S {d\hat{s}\over \hat{s}} f_{i/H_1}\left(x_1,\mu_f\right) f_{j/H_2}\left(x_2,\mu_f\right),
\end{displaymath}
$x_1= {\hat{s} / S}$ and $x_2= {s / \hat{s}}$. Here $S$ denotes the hadronic center-of-mass energy squared, and $s=x_1 x_2 S$ is the subprocess center-of-mass energy squared. The parameter $\mu_r$ denotes the (initial) renormalization scale and $\mu_f$ denotes the factorization scale. The choice of $\mu_r$ is arbitrary, which is only need to be in pQCD region ($\gg\Lambda_{\rm QCD}$) and usually people set its value as the typical momentum flow of the process; and for this process, $\mu_r$ is usually chosen as $m_t$. The function $f_{i/H_{\alpha}}(x_\alpha,\mu_f)$ ($\alpha=1$ or $2$) describes the probability of finding a parton of type $i$ with a light-front momentum fraction between $x_\alpha$ and $x_{\alpha} +dx_{\alpha}$ in the proton $H_{\alpha}$.

The partonic subprocess cross-section $\hat{\sigma}_{ij}$ up to NNLO level can be expanded as a power series of $\alpha_s$:
\begin{widetext}
\begin{eqnarray}
\hat{\sigma}_{ij} &=& \frac{1}{m_{t}^2} \left[f^0_{ij}(\rho,\mu_r,\mu_f) \alpha^2_s(\mu_r)+f^1_{ij}(\rho,\mu_r,\mu_f)\alpha^3_s(\mu_r) +f^2_{ij}(\rho,\mu_r,\mu_f)\alpha^4_s(\mu_r)+{\cal O}(\alpha_s^5)\right]
\end{eqnarray}
\end{widetext}
where $\rho=4m_{t}^2/s$. In the literature, the perturbative coefficients up to NNLO level have been calculated by various groups, e.g. Refs.\cite{Nason:1987xz, Nason:1989zy, Beenakker:1988bq, Beenakker:1990maa, Czakon:2008ii, Moch:2008qy, Beneke:2011mq, Kidonakis:2010dk, Baernreuther:2012ws, Czakon:2012pz, Czakon:2013goa}. The LO, NLO and NNLO coefficients $f^0_{ij}$, $f^1_{ij}$ and $f^2_{ij}$ can be explicitly read from the HATHOR program~\cite{Aliev:2010zk} and the Top++ program~\cite{Czakon:2011xx}, where $(ij)=\{(q\bar{q}),(gg),(gq),(g\bar{q})\}$ stands for the four production channels, respectively. By carefully identifying the $n_f$-terms specifically associated with the $\{\beta_i\}$-terms in $f^0_{ij}$, $f^1_{ij}$ and $f^2_{ij}$, and by using the degeneracy pattern of the renormalization group equation in a recursive way, one can determine the $\beta$ terms and thus the correct arguments of the strong couplings at each perturbative order. The remaining $n_f$ terms arise from quark loop contributions which are ultraviolet finite. A detailed determination of the PMC scales for $\hat{\sigma}_{ij}$ up to NNLO level, including a careful treatment of the separate renormalization scales of the Coulomb-type rescattering corrections appearing in the threshold region, have been presented in Refs.\cite{Brodsky:2012rj, Brodsky:2012sz}. We shall not repeat these formulae here; the interested readers may turn to those two references for details.

In doing the numerical analysis, we will first take the top-quark pole mass as $m_t=173.3$ GeV~\cite{toppole} and choose the parton distribution functions (PDF) as the CT14 version of the CTEQ collaboration~\cite{Dulat:2015mca}. The NNLO $\alpha_s$-running is adopted with its normalization fixed in $\overline {\rm MS}$-scheme using $\alpha_s(M_Z)=0.118$.

The setting of the factorization scale $\mu_f$ is a separate, important issue~\footnote{We have found that the factorization scale dependence is suppressed after applying the PMC~\cite{Wang:2014sua, Wang:2016wgw}; this can be explained by the fact that the pQCD series behaves much better after applying the PMC.}; however, a possible determination can be based on the light-front holographic QCD~\cite{Brodsky:2014yha}. It determines a scale $Q_0$ at the interface between nonpertubative and perturbative QCD. In the analysis given here, we will take $\mu_f=m_t$.

\begin{center}
\begin{table}[htb]
\begin{tabular}{|c||c|c|c|c|c|c|}
\hline
 & \multicolumn{3}{c|}{Conventional} & \multicolumn{3}{c|}{PMC} \\
 \hline
 ~~~$\mu_r$~~~& $m_t/2$  & $m_t$ & $2m_t$ & $m_t/2$ & $m_t$ & $2m_t$  \\
\hline
~$\sigma^{1.96\rm TeV}_{\rm Tevatron}$ & 7.54 & 7.29 & 7.01 & 7.43 & 7.43 & 7.43  \\
\hline
~$\sigma^{7\rm TeV}_{\rm LHC}$ & 172.07 & 167.67 & 160.46 & 174.97 & 174.98 & 174.99  \\
\hline
~$\sigma^{8\rm TeV}_{\rm LHC}$ & 244.87 & 239.03 & 228.94 & 249.16 & 249.18 & 249.19  \\
\hline
~$\sigma^{13\rm TeV}_{\rm LHC}$ & 792.36 & 777.72 & 746.92 & 807.80 & 807.83 & 807.86  \\
\hline
\end{tabular}
\caption{The NNLO top-quark pair production cross-sections for the Tevatron and LHC (in unit: pb), comparing conventional versus PMC scale settings. Here all production channels have been summed up. Three typical choices for the initial renormalization scales $\mu_{r}=m_t/2$, $m_t$ and $2m_t$ have been adopted. } \label{TevLHCtotcs}
\end{table}
\end{center}

We present the NNLO top-quark pair production cross-section at the hadronic colliders Tevatron and LHC for both conventional and PMC scale settings in Table~\ref{TevLHCtotcs}, where three typical initial renormalization scales are adopted. The results shown in Table~\ref{TevLHCtotcs} show that if one uses conventional scale-setting, the renormalization scale dependence of the NNLO cross-section is still about $6\% - 7\%$ for $\mu_r\in[m_t/2, 2m_t]$. If one analyzes the pQCD series in detail, one finds that the dependence of the NNLO cross-section on the guess of the renormalization scale using conventional scale-setting is due to cancelations among different orders, and the renormalization scale dependence of each perturbative term is rather large~\cite{Wang:2015lna}. Thus computing a finite number of additional higher-order terms could soften the renormalization scale dependence for the total cross-section to a certain degree, but it does not eliminate the dependence on the choice of the initial renormalization scale, especially when the detailed dependence on the renormalization scales at each order is also important.

When PMC scale-setting is used, the renormalization scales are fixed by using the renormalization group equation recursively, thus fixing the arguments of the strong couplings at each order. There is residual renormalization scale dependence due to unknown NNNLO and higher-order contributions, for example, we need to known the $\beta$-terms at the NNNLO level to fix the PMC scale of the NNLO-terms. Table~\ref{TevLHCtotcs} shows that the residual renormalization scale dependence of the NNLO total cross-section is negligibly small for $\mu_r\in[m_t/2, 2 m_t]$, which is less than $0.1\%$ even when taking a quite large initial renormalization scale range $\mu_r\in[m_t/4, 20m_t]$. The PMC scales are distinct at different orders, as in QED. Since the PMC scales are determined from perturbative input, any renormalization scale uncertainty of the pQCD series is transferred at finite order to the small uncertainty of the PMC scales.

If setting $\mu_r=m_t/2$ for conventional scale setting, the pQCD convergence is better than the cases of $\mu_r=m_t$ and $\mu_r=2m_t$, whose total cross-section is also close to the PMC prediction. Thus, for conventional scale setting, the best choice of an effective renormalization scale for top-quark pair production is $\mu_r\sim m_t/2$ other than the conventional suggested $m_t$~\cite{Wang:2014sua}. The choice of $\mu_r\sim m_t/2$ is also suggested in Ref.\cite{Czakon:2016dgf} by using the principle of fastest perturbative convergence.

\begin{figure}[htb]
\begin{center}
\includegraphics[width=0.45\textwidth]{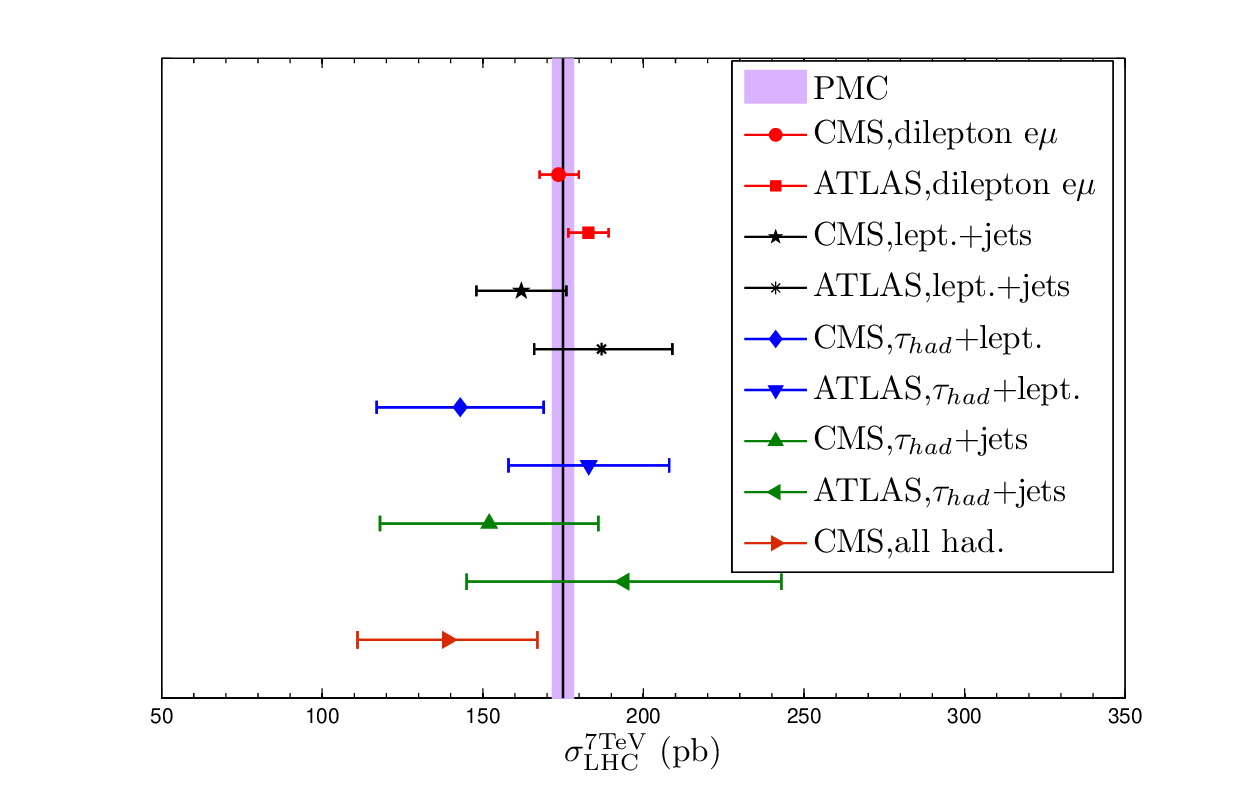}
\includegraphics[width=0.45\textwidth]{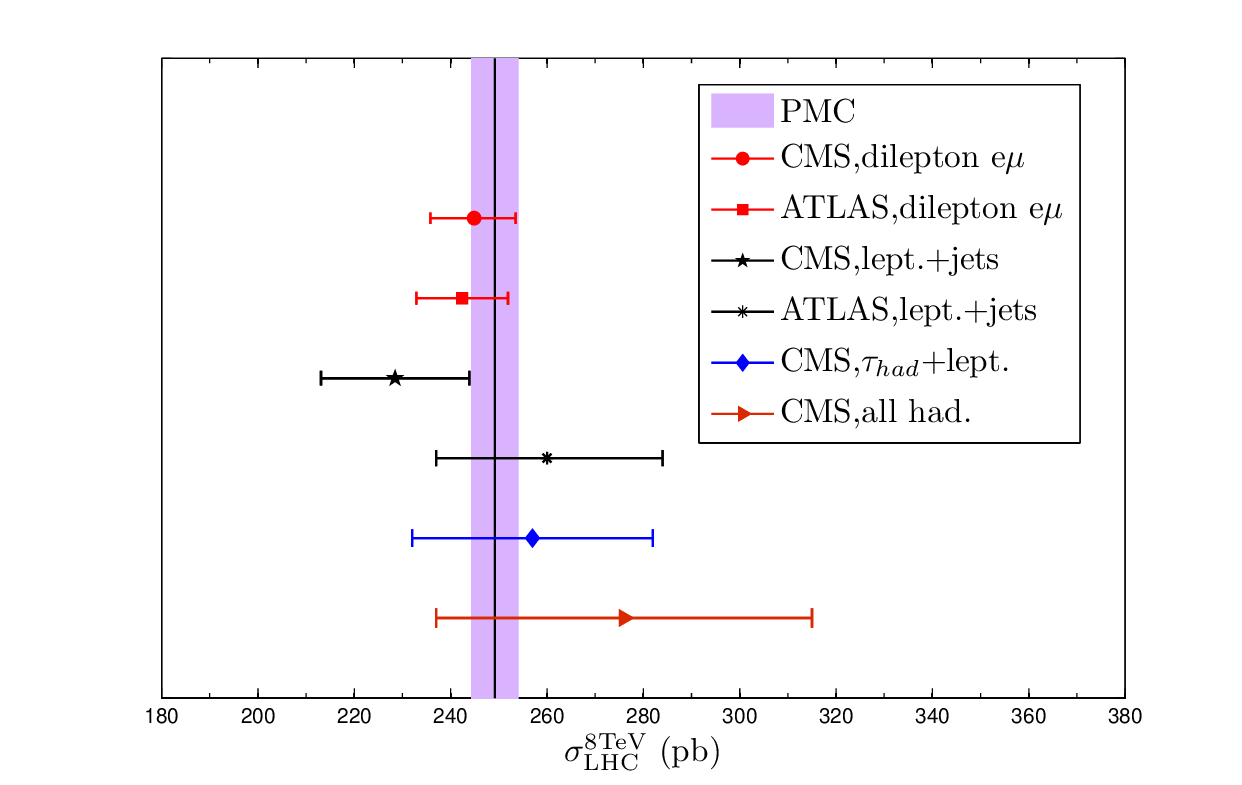}
\caption{Comparison of PMC prediction for the top-quark pair total cross-section with the LHC measurements for $\sqrt{S} = 7$ TeV(Up)~\cite{Chatrchyan:2013ual, Aad:2012vip, Chatrchyan:2013kff, Aad:2015dya, Chatrchyan:2012vs, Aad:2012qf, Chatrchyan:2016abc, Aad:2014kva, Khachatryan:2016mqs} and $\sqrt{S} = 8$ TeV(Down)~\cite{Khachatryan:2015fwh, Khachatryan:2014loa, Aad:2015pga, Chatrchyan:2016abc, Aad:2014kva, Khachatryan:2016mqs}. }
\label{totcslhc78TeV}
\end{center}
\end{figure}

Table~\ref{TevLHCtotcs} shows the PMC predictions for the top-quark pair total cross-section: $\sigma^{1.96\rm TeV}_{\rm Tevatron}=7.43^{+0.14}_{-0.13} ~\rm pb$ at the Tevatron, $\sigma^{7\rm TeV}_{\rm LHC}=175.0^{+3.5}_{-3.5} ~\rm pb$, $\sigma^{8\rm TeV}_{\rm LHC}=249.2^{+5.0}_{-4.9} ~\rm pb$, and $\sigma^{13\rm TeV}_{\rm LHC}=807.8^{+16.0}_{-15.8} ~\rm pb$ at the LHC for $\sqrt{S}=7$, 8 and 13 TeV, respectively. Those predictions agree with the Tevatron and LHC measurements~\cite{Aaltonen:2013wca, Chatrchyan:2013ual, Aad:2012vip, Chatrchyan:2013kff, Aad:2015dya, Chatrchyan:2012vs, Aad:2012qf, Chatrchyan:2016abc, Aad:2014kva, Khachatryan:2016mqs, Khachatryan:2015fwh, Khachatryan:2014loa, Aad:2015pga, Khachatryan:2016kzg, Khachatryan:2015uqb, Aaboud:2015AAAA, Aaboud:2016pbd}. A comparison of the PMC prediction for the top-quark pair production cross-section with the LHC measurements is shown in Fig.(\ref{totcslhc78TeV}) for $\sqrt{S} = 7$ TeV and $8$ TeV. As in Ref.\cite{Khachatryan:2016mqs}, the theoretical error bands in Fig.(\ref{totcslhc78TeV}) is estimated by using the CT14 error PDF sets~\cite{Dulat:2015mca} with the range of $\alpha_s(M_Z)\in [0.117, 0.119]$.

It is important to study the ratio of total cross sections $R^{8/7}=(\sigma^{8\rm TeV}_{\rm LHC})/(\sigma^{7\rm TeV}_{\rm LHC})$, since the experimental uncertainties, which are correlated between the two analyses (at $\sqrt{S}$ = 7 or 8 TeV) cancel out, leading to an improved precision in comparison to the individual measurements. The predicted cross-section ratio by the PMC is $R^{8/7}|_{\rm PMC}=1.42\pm0.04$, which shows excellent agreement with the latest CMS measurement $R^{8/7}|_{\rm CMS}= 1.43\pm0.04 \pm0.07 \pm0.05$~\cite{Chatrchyan:2016abc}.

As we have shown above, the PMC provides a comprehensive and self-consistent pQCD explanation for the top-quark pair production cross-section as well as the top-quark pair forward-backward asymmetry. The behavior of the top-quark pair production cross section allows a direct determination of the top-quark pole mass by comparing the pQCD prediction with the data.

Following the method of Ref.\cite{Aaboud:2016000}, we define a likelihood function
\begin{eqnarray}
f(m_t)&=&\int^{+\infty}_{-\infty} f_{\rm th}(\sigma|m_t)\cdot f_{\rm exp}(\sigma|m_t)\; d\sigma.
\label{likelifunction}
\end{eqnarray}
Here $f_{\rm th}(\sigma|m_t)$ is the normalized Gaussian distribution, which is defined as
\begin{eqnarray}
f_{\rm th}(\sigma|m_t) = \frac{1}{\sqrt{2\pi}\Delta \sigma_{\rm th}(m_t)}
 \exp \left[-\frac{\left(\sigma-\sigma_{\rm th}(m_t)\right)^2}
 {2 \Delta\sigma^2_{\rm th}(m_t)}\right].
\end{eqnarray}
The top-quark pair production cross-section is a function of the top-quark pole mass $m_t$, decreases with increasing $m_t$.  It can be parameterized as~\cite{Beneke:2011mq}
\begin{widetext}
\begin{eqnarray}
\sigma_{\rm th}(m_t)&=&\left(\frac{172.5}{m_t/{\rm GeV}}\right)^4\left(c_0+c_1(\frac{m_t}{\rm GeV}-172.5) +c_2  \times(\frac{m_t}{\rm GeV}-172.5)^2+c_3(\frac{m_t}{\rm GeV}-172.5)^3\right),
\label{massdepdPMC}
\end{eqnarray}
\end{widetext}
where all masses are given in units of GeV. $\Delta\sigma_{\rm th}(m_t)$ stands for the maximum error of the cross-section for a fixed $m_t$; it is estimated by using the CT14 error PDF sets~\cite{Dulat:2015mca} with range of $\alpha_s(M_Z)\in [0.117, 0.119]$. The determined coefficients $c_{0,1,2,3}$ are given in Table~\ref{tab_coeffi}.

\begin{widetext}
\begin{center}
\begin{table}[htb]
\begin{tabular}{|c|c|c|c|c|c|}
\hline
\multicolumn{2}{|c|}{~~~~} &~~~$c_0~[\rm pb]$~~~  &~~~$c_1~[\rm pb]$~~~  &~~~$c_2~[\rm pb]$~~~ &~~~$c_3~[\rm pb]$~~~ \\
\hline
& $\sigma_{\rm th}$                                   & $7.6181$ & $-0.06140$ & $2.1135\times10^{-4}$ & $-1.9319\times10^{-6}$ \\
Tevatron& $\sigma_{\rm th}+\Delta\sigma^{+}_{\rm th}$ & $7.7580$ & $-0.06261$ & $2.1711\times10^{-4}$ & $-1.9923\times10^{-6}$ \\
& $\sigma_{\rm th}-\Delta\sigma^{-}_{\rm th}$         & $7.4796$ & $-0.06019$ & $2.0572\times10^{-4}$ & $-1.8750\times10^{-6}$ \\
\hline
& $\sigma_{\rm th}$                                            & $179.2422$ & $-1.2311$ & $ 4.7155\times10^{-3}$ & $-3.3920\times10^{-5}$ \\
LHC$|_{7\rm TeV}$& $\sigma_{\rm th}+\Delta\sigma^{+}_{\rm th}$ & $182.8195$ & $-1.2590$ & $ 4.8479\times10^{-3}$ & $-3.5338\times10^{-5}$ \\
& $\sigma_{\rm th}-\Delta\sigma^{-}_{\rm th}$                  & $175.7093$ & $-1.2037$ & $ 4.5866\times10^{-3}$ & $-3.2489\times10^{-5}$ \\
\hline
& $\sigma_{\rm th}$                                            & $255.0975$ & $-1.5718$ & $ 5.2644\times10^{-3}$ & $-4.2394\times10^{-5}$ \\
LHC$|_{8\rm TeV}$& $\sigma_{\rm th}+\Delta\sigma^{+}_{\rm th}$ & $260.1779$ & $-1.6078$ & $ 5.4191\times10^{-3}$ & $-4.4240\times10^{-5}$ \\
& $\sigma_{\rm th}-\Delta\sigma^{-}_{\rm th}$                  & $250.0801$ & $-1.5364$ & $ 5.1128\times10^{-3}$ & $-4.0618\times10^{-5}$ \\
\hline
& $\sigma_{\rm th}$                                             & $825.5955$ & $-3.2873$ & $5.1997\times10^{-3}$ & $-1.0274\times10^{-4}$ \\
LHC$|_{13\rm TeV}$& $\sigma_{\rm th}+\Delta\sigma^{+}_{\rm th}$ & $841.9260$ & $-3.3675$ & $5.4202\times10^{-3}$ & $-1.0730\times10^{-4}$ \\
& $\sigma_{\rm th}-\Delta\sigma^{-}_{\rm th}$                   & $809.4638$ & $-3.2084$ & $4.9863\times10^{-3}$ & $-9.8738\times10^{-5}$ \\
\hline
\end{tabular}
\caption{The coefficients $c_{0,1,2,3}$ as determined from the PMC predictions for the top-quark pair cross-section by varying the top-quark pole mass from 160 GeV to 190 GeV. The notation $[\sigma_{\rm th}(m_t) + \Delta\sigma^{+}_{\rm th}(m_t)]$ indicates that the coefficients are determined using the maximum cross section within its allowable parameter range, and $[\sigma_{\rm th}(m_t)-\Delta\sigma^{-}_{\rm th}(m_t)]$ corresponds to the minimum cross section. }
\label{tab_coeffi}
\end{table}
\end{center}
\end{widetext}

In order to determine the precise values for the coefficients $c_{0,1,2,3}$, we have used a wide range of the top-quark pole mass, i.e. $m_t\in [{\rm 160\; GeV}, {\rm 190\; GeV}]$. We define $\sigma_{\rm th}(m_t)$ as the cross-section at a fixed $m_t$, where all input parameters are taken at their central values, $[\sigma_{\rm th}(m_t) + \Delta\sigma^{+}_{\rm th}(m_t)]$ is the maximum cross-section within the allowable parameter range, and $[\sigma_{\rm th}(m_t)-\Delta\sigma^{-}_{\rm th}(m_t)]$ is the minimum value. Similarly, $f_{\rm exp}(\sigma|m_t)$ is the normalized Gaussian distribution
\begin{eqnarray}
f_{\rm exp}(\sigma|m_t) = \frac{1}{\sqrt{2\pi}\Delta \sigma_{\rm exp}(m_t)} \exp{ \left[-\frac{\left(\sigma-\sigma_{\rm exp}(m_t)\right)^2}
 {2\Delta\sigma^2_{\rm exp}(m_t)} \right]},
\end{eqnarray}
where $\sigma_{\rm exp}(m_t)$ is the measured cross-section, and $\Delta\sigma_{\rm exp}(m_t)$ is the uncertainty for $\sigma_{\rm exp}(m_t)$.

\begin{figure}[htb]
\centering
\includegraphics[width=0.45\textwidth]{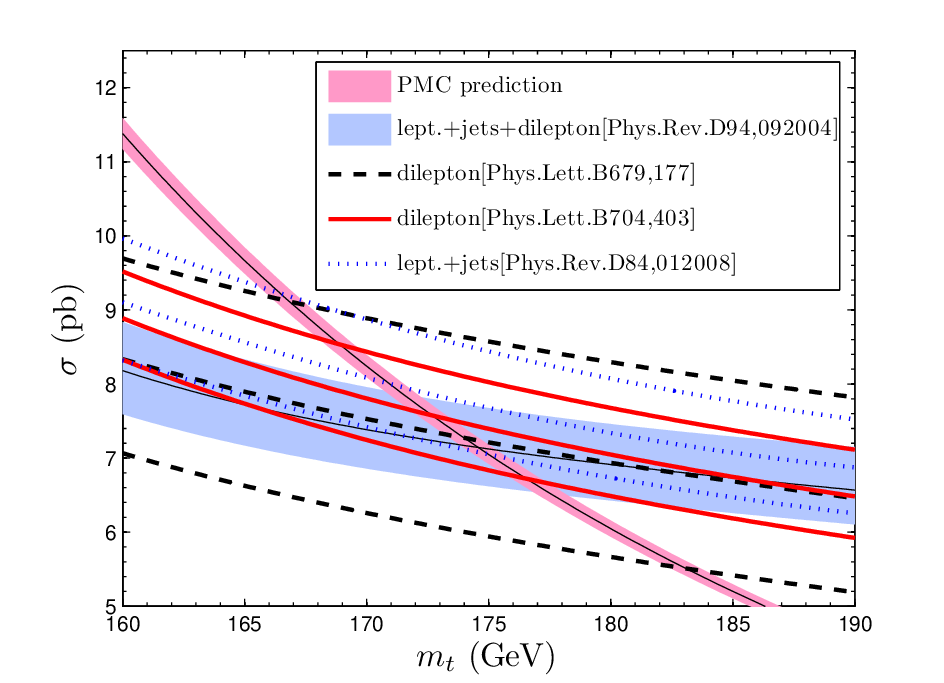}
\caption{The top-quark pair production cross-section using PMC scale-setting versus the top-quark pole mass at the Tevatron with the collision energy $\sqrt{S}=1.96$ TeV. As for the two shaded bands, the thinner one and the thicker one are for the PMC prediction and the combined experimental result from Ref.\cite{Abazov:2016ekt}, respectively. The dashed, solid, and dotted lines are measurements for the dilepton channel~\cite{Abazov:2009si, Abazov:2011cq} and the lepton + jets channel~\cite{Abazov:2011mi}, respectively. The upper and lower lines indicate the error range of the corresponding measurements.}
\label{figureTOPde196}
\end{figure}

\begin{figure}[htb]
\centering
\includegraphics[width=0.45\textwidth]{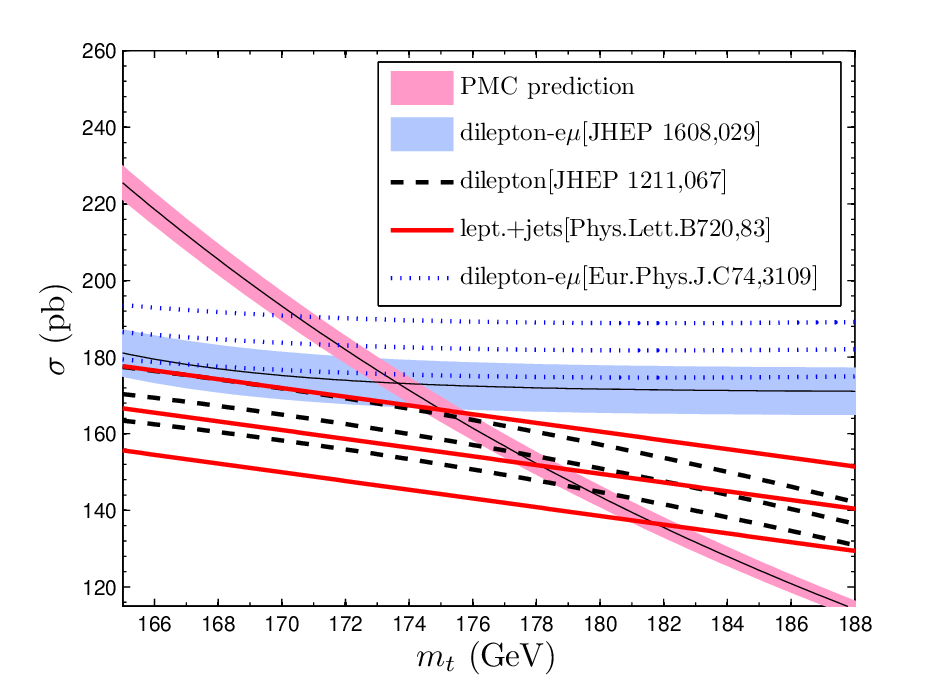}
\includegraphics[width=0.45\textwidth]{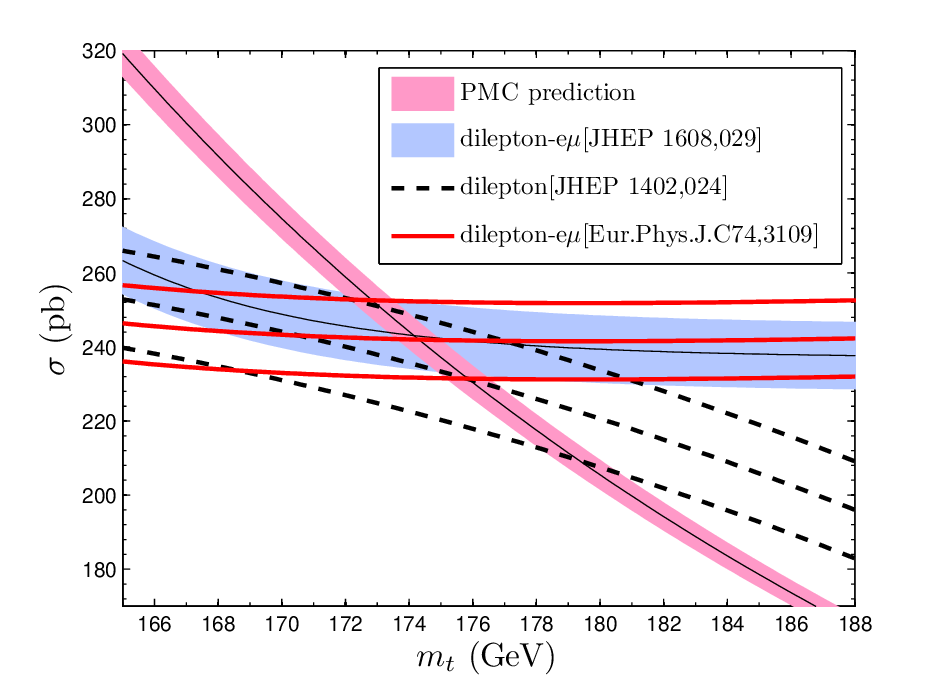}
\caption{The top-quark pair production cross-section using PMC scale-setting versus the top-quark pole mass at the LHC with the collision energy $\sqrt{S}=7$ TeV (Up) and $\sqrt{S}=8$ TeV (Down), respectively. As for the two shaded bands, the thinner one and the thicker one are for the PMC prediction and the combined experimental result from Ref.\cite{Khachatryan:2016mqs}, respectively. In the upper diagram, the dashed, the solid and the dotted lines are measurements for the dilepton~\cite{Chatrchyan:2012bra} and the lepton + jets~\cite{Chatrchyan:2012ria}, and the dilepton-e$\mu$~\cite{Aad:2014kva} channels, respectively. In the lower diagram, the dashed and the solid lines are measurements for the dilepton~\cite{Chatrchyan:2013faa} and the dilepton-e$\mu$~\cite{Aad:2014kva} channels, respectively. The upper and lower lines indicate the error range of the corresponding measurements. }
\label{figureTOPde7}
\end{figure}

We present the top-quark pair NNLO production cross-section (\ref{massdepdPMC}) versus the top-quark pole mass at different hadron-hadron collision energies in Figs.(\ref{figureTOPde196}, \ref{figureTOPde7}). The coefficients $c_{0,1,2,3}$ are determined by the PMC predictions. In these figures, the experimental measurements are presented for comparison, where the thinnest shaded bands are for the PMC predictions and the thickest shaded bands are for the combined experimental results respectively. The agreement of the PMC predictions with the measurements, as shown by Figs.(\ref{figureTOPde196}, \ref{figureTOPde7}), makes it possible to achieve reliable predictions for top-quark pole mass. A precise range of values for the pole mass can thus be achieved in comparison with pQCD predictions based on conventional scale-setting. In the following, we will determine the top-quark pole mass such that the maximum value of the likelihood function (\ref{likelifunction}) is achieved.

The D0 collaboration determined the top-quark pole mass by comparing the theoretical predictions based on conventional scale-setting with the measurements of the top-quark pair production cross-sections at the Tevatron~\cite{Abazov:2009si, Abazov:2011cq, Abazov:2011mi, Abazov:2016ekt}. The results for various production channels are presented in Table~\ref{TevatronMasscom}. As a comparison, we present our predictions using PMC scale-setting in Table~\ref{TevatronMasscom}. For the calculation of the likelihood function (\ref{likelifunction}), we have used the experimental measurements in these references as the input for $f_{\rm exp}(\sigma|m_t)$.

\begin{widetext}
\begin{center}
\begin{table}[htb]
\begin{tabular}{|c|c|c|c|c|}
\hline
~~ ~~ & \multicolumn{2}{|c|}{dilepton} & lept.+jets & lept.+jets+dilepton  \\
\hline
Conv. & $171.5^{+9.9}_{-8.8}$ ~\cite{Abazov:2009si} & $171.6\pm4.3$ ~\cite{Abazov:2011cq, Beneke:2012wb} & $166.7^{+5.2}_{-4.5}$ ~\cite{Abazov:2011mi, Abazov:2011pta} & $172.8^{+3.4}_{-3.2}$ ~\cite{Abazov:2016ekt} \\
\hline
PMC & $174.0^{+8.5}_{-9.8}$ & $172.7^{+4.1}_{-4.3}$ & $171.1\pm4.9$ & $174.6^{+3.1}_{-3.2}$ \\
\hline
\end{tabular}
\caption{Top-quark pole mass (in unit GeV) determined by D0 collaboration~\cite{Abazov:2009si, Abazov:2011cq, Abazov:2011mi, Abazov:2016ekt}, where the theoretical predictions for top-quark pair production is based on conventional (Conv.) scale-setting. Our predictions using PMC scale-setting are presented as a comparison. }
\label{TevatronMasscom}
\end{table}
\end{center}
\end{widetext}

Table~\ref{TevatronMasscom} shows that the top-quark pole mass determined from the dilepton channel which are measured at the Tevatron Run I stage possesses the largest uncertainty~\cite{Abazov:2009si}. It will be improved by more precise data for the dilepton and the lepton + jets channels obtained at the Run II stage~\cite{Abazov:2011cq, Abazov:2011mi, Abazov:2016ekt}.

\begin{figure}[htb]
\centering
\includegraphics[width=0.45\textwidth]{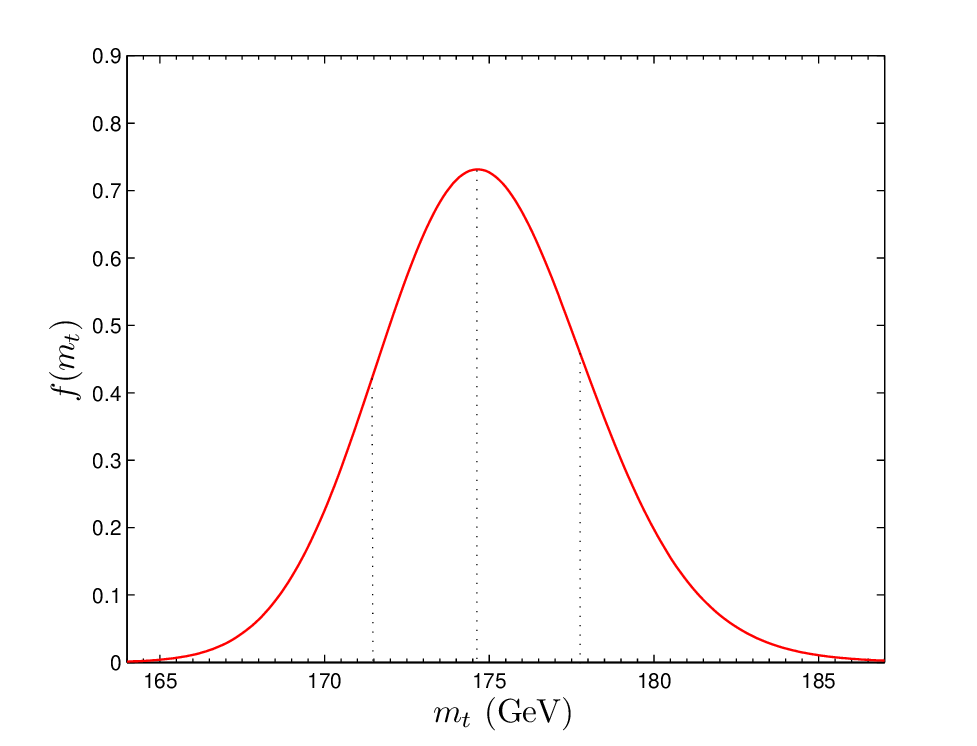}
\caption{The likelihood function $f(m_t)$ at the Tevatron obtained by using the measured combined inclusive top-quark pair cross-section of Ref.\cite{Abazov:2016ekt} as the experimental input. The three vertical dotted lines indicate the maximum of $f(m_t)$ and the edges of the $68\%$ area of the maximum of $f(m_t)$. }
\label{lifunction196TeV}
\end{figure}

We present the likelihood function defined in Eq.(\ref{likelifunction}) at the Tevatron in Fig.(\ref{lifunction196TeV}), where the measured combined inclusive top-quark pair cross-section of Ref.\cite{Abazov:2016ekt} are adopted as the experimental input. By evaluating the likelihood function, we obtain $m_t=174.6^{+3.1}_{-3.2}$ GeV, where the central value is extracted from the maximum of the likelihood function, and the error ranges are obtained from the $68\%$ area around the maximum. As indicated by Figs.(\ref{figureTOPde196}, \ref{figureTOPde7}), due to the elimination of renormalization scale uncertainty. The PMC predictions have less uncertainty compared to the predictions by using conventional scale-setting. Thus the uncertainty of the precision of top-quark pole mass is dominated by the experimental errors. For example, the PMC determination for the pole mass via the combined dilepton and the lepton + jets channels data is about $1.8\%$, which is almost the same as that of the recent determination by the D0 collaboration, $172.8^{+3.4}_{-3.2}$ GeV~\cite{Abazov:2016ekt} whose error is $\sim 1.9\%$.

\begin{table}[htb]
\begin{center}
\begin{tabular}{|c|c|c|c|}
\hline
~~ ~~ & dilepton  & \multicolumn{2}{|c|}{dilepton-e$\mu$}  \\
\hline
Conv. & $177.0^{+3.6}_{-3.3}$~\cite{Chatrchyan:2012bra, Chatrchyan:2013haa} & $171.4\pm2.6$~\cite{Aad:2014kva} & $174.1^{+2.2}_{-2.4}$~\cite{Khachatryan:2016mqs}  \\
\hline
PMC & $177.5\pm2.4$ & $171.8\pm1.6$ & $173.7\pm1.5$ \\
\hline
\end{tabular}
\caption{Top-quark pole mass (in unit GeV) determined by CMS and ATLAS collaborations at $\sqrt{S}=7$ TeV~\cite{Chatrchyan:2012bra, Chatrchyan:2013haa, Aad:2014kva, Khachatryan:2016mqs}, where the theoretical predictions for top-quark pair production is based on conventional (Conv.) scale-setting. Our predictions using PMC scale-setting are presented as a comparison. }
\label{LHC7TeVMasscom}
\end{center}
\end{table}

\begin{table}[htb]
\begin{center}
\begin{tabular}{|c|c|c|c|}
\hline
~~ ~~ & \multicolumn{2}{|c|}{dilepton-e$\mu$}  \\
\hline
Conv. &  $174.1\pm 2.6$~\cite{Aad:2014kva} & $174.6^{+2.3}_{-2.5}$~\cite{Khachatryan:2016mqs}  \\
\hline
PMC & $174.3\pm1.7$ & $174.2\pm1.7$  \\
\hline
\end{tabular}
\caption{Top-quark pole mass (in unit GeV) determined by CMS and ATLAS collaborations at $\sqrt{S}=8$ TeV~\cite{Aad:2014kva, Khachatryan:2016mqs}, where the theoretical predictions for top-quark pair production is based on conventional (Conv.) scale-setting. Our predictions using PMC scale-setting are presented as a comparison. }
\label{LHC8TeVMasscom}
\end{center}
\end{table}

The CMS and ATLAS collaborations have determined the top-quark pole mass by using measurements of top-quark pair production cross-sections at the LHC~\cite{Chatrchyan:2012bra, Chatrchyan:2013haa, Aad:2014kva, Khachatryan:2016mqs} together with the theoretical predictions derived from conventional scale-setting; the results for various production channels are presented in Tables~\ref{LHC7TeVMasscom} and \ref{LHC8TeVMasscom} for $\sqrt{S}=7$ and 8 TeV, respectively. As a comparison, we also present our predictions using PMC scale-setting in the two Tables. Similarly, for calculating the likelihood function (\ref{likelifunction}), we use the experimental measurements in those references as the input for $f_{\rm exp}(\sigma|m_t)$.

\begin{figure}[htb]
\centering
\includegraphics[width=0.45\textwidth]{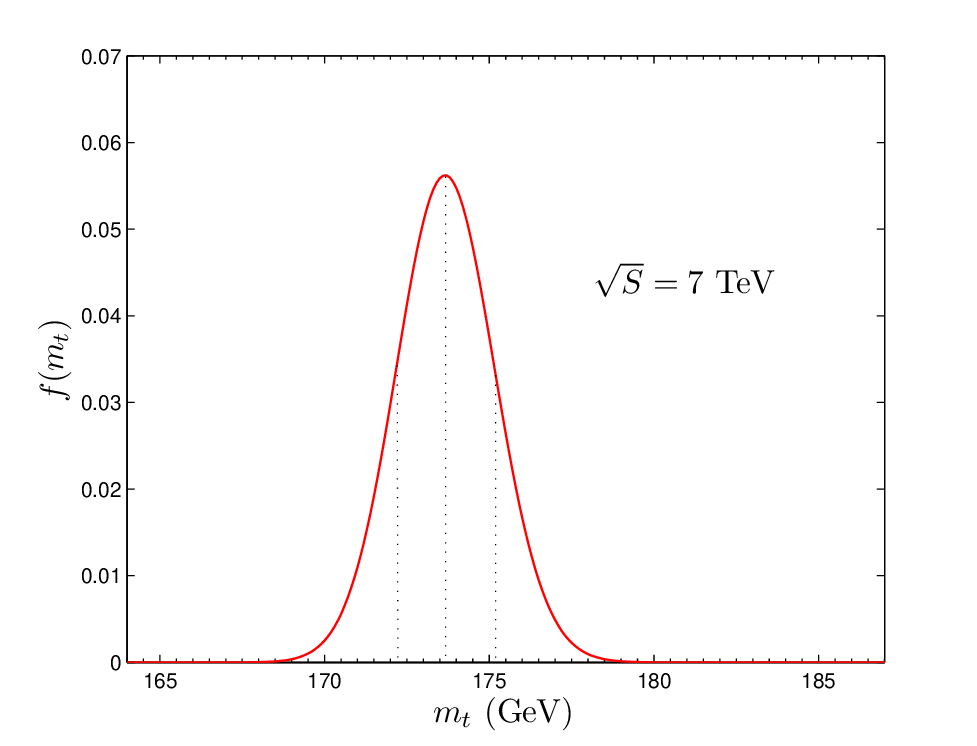}
\includegraphics[width=0.45\textwidth]{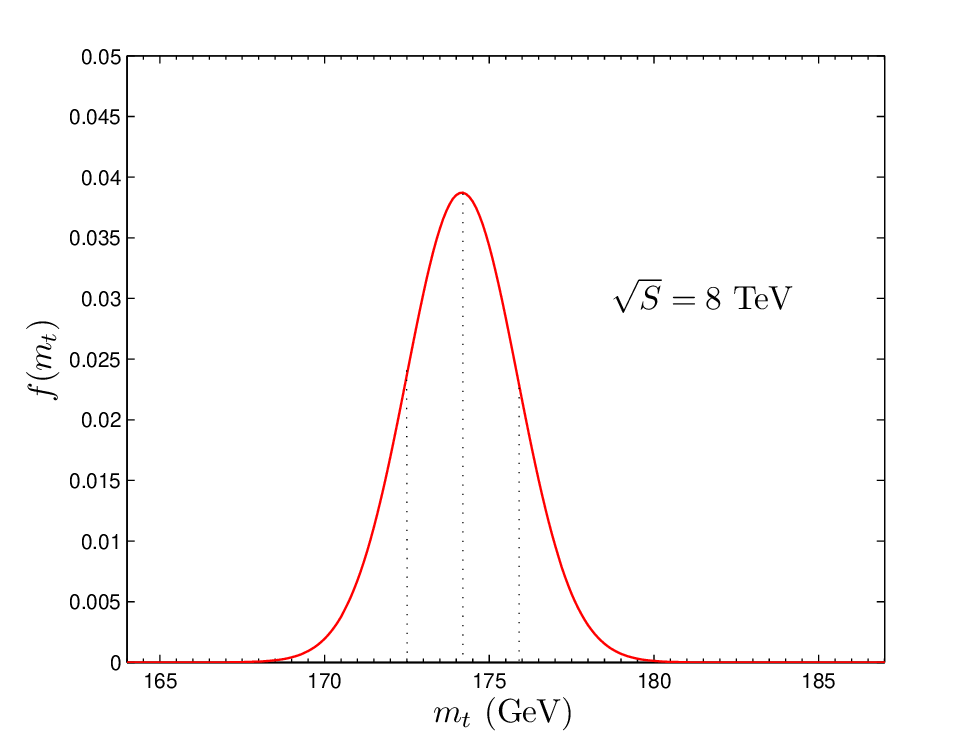}
\caption{The likelihood function $f(m_t)$ defined in Eq.(\ref{likelifunction}) at the LHC with $\sqrt{S}=$7 TeV (Up) and $\sqrt{S}=$8 TeV (Down). The three vertical dotted lines stand for the maximum of $f(m_t)$ and for the $68\%$ area around the maximum of $f(m_t)$. }
\label{lifunction78TeV}
\end{figure}

By using the measured cross section $\sigma_{\rm exp}(m_t)$ together with its error $\Delta\sigma_{\rm exp}(m_t)$ from the latest CMS measurement~\cite{Khachatryan:2016mqs}, we present the likelihood functions at the LHC in Fig.(\ref{lifunction78TeV}). Because the experimental uncertainty at the LHC is smaller than that of Tevatron, the determined top-quark pole mass by using the LHC data has better precision in comparison with the analysis using the Tevatron data. By evaluating the likelihood functions, we obtain $m_t=173.7\pm1.5$ GeV for $\sqrt{S}=7$ TeV, and $m_t=174.2\pm1.7$ GeV for $\sqrt{S}=8$ TeV. The precision of the top-quark pole masses determination is improved to be (${\pm1.5}$) for $\sqrt{S}=7$ TeV and (${\pm1.7}$) for $\sqrt{S}=8$ TeV.

By evaluating the likelihood function (\ref{likelifunction}) using the corresponding measurements of the latest Tevatron and LHC collaborations, we obtain the following predictions for the top-quark pole mass,
\begin{eqnarray}
m_t|_{{\rm Tevatron},\sqrt{S} = 1.96 {\rm TeV}} &=& 174.6^{+3.1}_{-3.2} ~{\rm GeV}, \\
m_t|_{{\rm LHC},\sqrt{S} = 7 {\rm TeV}} &=& 173.7^{+1.5}_{-1.5} ~\rm GeV, \\
m_t|_{{\rm LHC},\sqrt{S} = 8 {\rm TeV}} &=& 174.2^{+1.7}_{-1.7} ~\rm GeV.
\end{eqnarray}
By using the relation between the pole mass and the $\overline{\rm MS}$ mass up to four-loop level~\cite{Marquard:2015qpa, Kataev:2015gvt}, we can convert the top-quark pole mass to the $\overline{\rm MS}$ definition. For $\mu_r=m_t$, we obtain
\begin{eqnarray}
m^{\overline{\rm MS}}_t(m_t)|_{{\rm Tevatron},\sqrt{S} = 1.96 {\rm TeV}} &=& 164.0^{+2.9}_{-3.0} ~{\rm GeV}, \\
m^{\overline{\rm MS}}_t(m_t)|_{{\rm LHC},\sqrt{S} = 7 {\rm TeV}} &=& 163.1^{+1.4}_{-1.4} ~\rm GeV, \\
m^{\overline{\rm MS}}_t(m_t)|_{{\rm LHC},\sqrt{S} = 8 {\rm TeV}} &=& 163.6^{+1.6}_{-1.6} ~\rm GeV.
\end{eqnarray}
The weighted average of those predictions then leads to
\begin{equation}
m_t=174.0\pm1.1 \; {\rm GeV} \;\;{\rm and}\;\; m^{\overline{\rm MS}}_t(m_t)=163.4\pm1.0 \; {\rm GeV}.
\end{equation}

\begin{figure}[htb]
\centering
\includegraphics[width=0.45\textwidth]{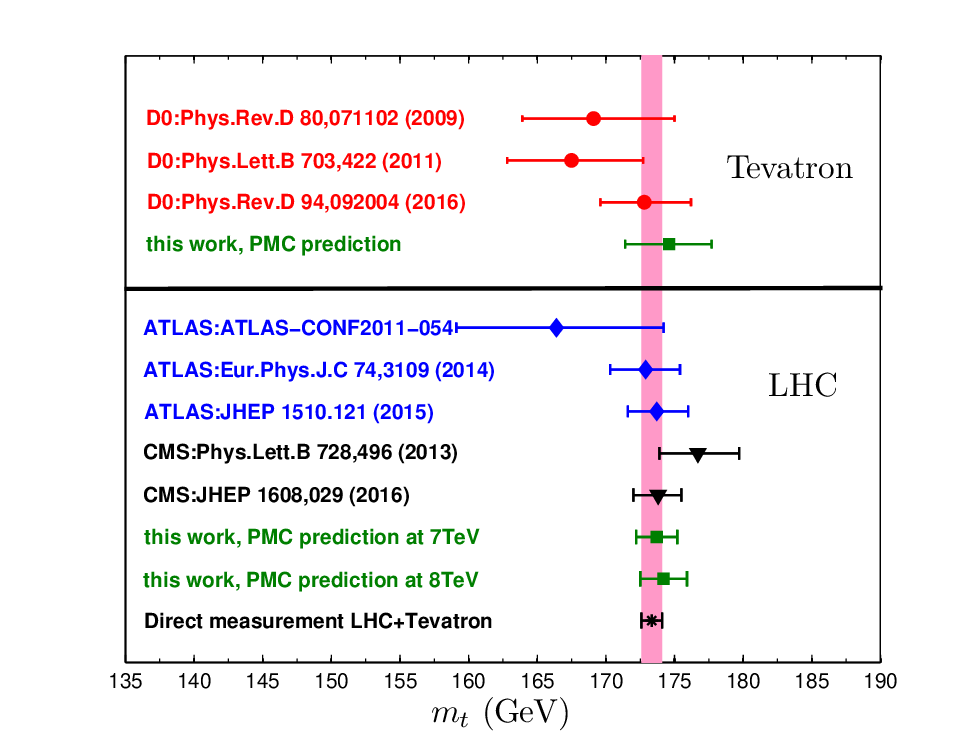}
\caption{A summary of the top-quark pole mass determined indirectly from the top-quark pair production channels at the Tevatron and LHC. As a reference, the combination of Tevatron and LHC direct measurements of the top-quark mass is presented as a shaded band. It gives $m_t=173.34\pm0.76$ GeV~\cite{ATLAS:2014wva}. }
\label{toppolesummary}
\end{figure}

We summarize the top-quark pole masses determined at both the Tevatron and LHC in Fig.(\ref{toppolesummary}), where our PMC predictions and previous predictions from other collaborations~\cite{Abazov:2009ae, Abazov:2011pta, Abazov:2016ekt, Aaboud:2016000, Aad:2014kva, Aad:2015waa, Chatrchyan:2013haa, Khachatryan:2016mqs} are presented. For reference, the combination of Tevatron and LHC direct measurements of the top-quark mass is presented as a shaded band, giving $m_t=173.34\pm0.76$ GeV~\cite{ATLAS:2014wva}. It shows that our new top-quark pole mass determined by PMC agree with the combination of Tevatron and LHC direct measurements.

\section{Summary}

We have achieved precise predictions for the top-quark pair production cross-section with minimal dependence on the choice of the initial renormalization scale by using PMC. The resulting predictions are in agreement with measurements done by both the Tevatron and the LHC Collaborations. We have given a new determination of the top-quark pole mass by comparing the PMC prediction for the top-quark pair cross-sections with the latest measurements; a detailed comparison of previous determinations given in the literature has also been presented. Our new determination of the top-quark pole masses provide complementary information compared to direct measurements.

The determined top-quark pole masses are cross-checked by other determinations used different techniques. Typically, the mass $m_t=175.8^{+2.7}_{-2.4}$ GeV from an electroweak fits~\cite{Baak:2012kk}, the mass $m_t=173.5\pm3.0\pm0.9$ GeV reconstructed from lepton + $J/\psi$ from b-jet~\cite{Khachatryan:2016pek}, the mass $m_t=173.2\pm1.6$ GeV from dilepton kinematic distributions\cite{2017mlepton} and the best direct measurement results $m_t=172.99\pm0.85$ GeV from ATLAS~\cite{Aaboud:2016igd} and $m_t=172.35\pm0.51$ GeV from CMS~\cite{Khachatryan:2015hba}. The consistency of the pQCD predictions using the PMC with all of the collider measurements at different energies and different techniques provides an important verification of QCD.

The PMC provides a systematic, rigorous method for eliminating renormalization scheme-and-scale ambiguities at each order in perturbation theory. As we have shown in our previous papers, the PMC is applicable to a wide variety of perturbatively calculable processes. In each case, the {\it ad hoc} renormalization scale uncertainty conventionally assigned to the pQCD predictions can be eliminated. The residual renormalization scale dependence due to uncalculated high-order terms are usually small due to a more convergent pQCD series. The PMC, with its solid physical and rigorous theoretical background, thus will greatly improve the precision of tests of the Standard Model.

\hspace{1cm}

{\bf Acknowledgements}: The authors would like to thank Hui-Lan Liu for helpful discussions. This work was supported in part by the Natural Science Foundation of China under Grant No.11547010, No.11625520, No.11705033 and No.11325525; by the Project of Guizhou Provincial Department of Science and Technology under Grant No.2016GZ42963 and the Key Project for Innovation Research Groups of Guizhou Provincial Department of Education under Grant No.KY[2016]028 and No.KY[2017]067; and by the Department of Energy Contract No.DE-AC02-76SF00515. SLAC-PUB-16934.


\begin{thebibliography}{99}

%\cite{Dalitz:1992np}
\bibitem{Dalitz:1992np}
  R.~H.~Dalitz and G.~R.~Goldstein,
  %Analysis of top-antitop production and dilepton decay events and the top quark mass,
  Phys.\ Lett.\ B {\bf 287}, 225 (1992).

%\cite{Chatrchyan:2013haa}
\bibitem{Chatrchyan:2013haa}
  S.~Chatrchyan {\it et al.} [CMS Collaboration],
  %Determination of the top-quark pole mass and strong coupling constant from the t t-bar production cross section in pp collisions at $\sqrt{s}$ = 7 TeV,
  Phys.\ Lett.\ B {\bf 728}, 496 (2014)
  Erratum: [Phys.\ Lett.\ B {\bf 738}, 526 (2014)].
  %[arXiv:1307.1907 [hep-ex]].

%\cite{Aaboud:2016000}
\bibitem{Aaboud:2016000}
  M.~Aaboud {\it et al.} [ATLAS Collaboration],
  %Determination of the Top-Quark Mass from the $t\bar{t}$ Cross Section Measurement in pp Collisions at $\sqrt{S}=7$ TeV with the ATLAS detector,
  ATLAS-CONF-2011-054.

%\cite{ATLAS:2014wva}
\bibitem{ATLAS:2014wva}
  [ATLAS and CDF and CMS and D0 Collaborations],
  %First combination of Tevatron and LHC measurements of the top-quark mass,
  arXiv:1403.4427 [hep-ex].

%\cite{Buckley:2011ms}
\bibitem{Buckley:2011ms}
  A.~Buckley {\it et al.},
  %General-purpose event generators for LHC physics,
  Phys.\ Rept.\  {\bf 504}, 145 (2011).
  %[arXiv:1101.2599 [hep-ph]].

%\cite{Fleming:2007qr}
\bibitem{Fleming:2007qr}
  S.~Fleming, A.~H.~Hoang, S.~Mantry and I.~W.~Stewart,
  %Jets from massive unstable particles: Top-mass determination,
  Phys.\ Rev.\ D {\bf 77}, 074010 (2008).
  %[hep-ph/0703207].

%\cite{Khachatryan:2016mqs}
\bibitem{Khachatryan:2016mqs}
  V.~Khachatryan {\it et al.} [CMS Collaboration],
  %Measurement of the t-tbar production cross section in the e-mu channel in proton-proton collisions at sqrt(s) = 7 and 8 TeV,
  JHEP {\bf 1608}, 029 (2016).
  %[arXiv:1603.02303 [hep-ex]].

%\cite{Tarrach:1980up}
\bibitem{Tarrach:1980up}
  R.~Tarrach,
  %The Pole Mass in Perturbative QCD,
  Nucl.\ Phys.\ B {\bf 183}, 384 (1981).

\bibitem{Kronfeld:1998di}
  A.~S.~Kronfeld,
  %The Perturbative pole mass in QCD,
  Phys.\ Rev.\ D {\bf 58}, 051501 (1998).

%\cite{Abazov:2016ekt}
\bibitem{Abazov:2016ekt}
  V.~M.~Abazov {\it et al.} [D0 Collaboration],
  %Measurement of the inclusive $t\bar{t}$ production cross section in $p\bar{p}$ collisions at $\sqrt{s}=1.96$ TeV and determination of the top quark pole mass,
  Phys.\ Rev.\ D {\bf 94}, 092004 (2016).
  %[arXiv:1605.06168 [hep-ex]].

%\cite{Aad:2014kva}
\bibitem{Aad:2014kva}
  G.~Aad {\it et al.} [ATLAS Collaboration],
  %Measurement of the $t\overline{t}$ production cross-section using $e\mu $ events with $b$ -tagged jets in $pp$ collisions at $\sqrt{s}=7$ and 8 TeV with the ATLAS detector,
  Eur.\ Phys.\ J.\ C {\bf 74}, 3109 (2014).
  %[arXiv:1406.5375 [hep-ex]].

%\cite{Wu:2013ei}
\bibitem{Wu:2013ei}
  X.~G.~Wu, S.~J.~Brodsky and M.~Mojaza,
  %The Renormalization Scale-Setting Problem in QCD,
  Prog.\ Part.\ Nucl.\ Phys.\  {\bf 72}, 44 (2013).

%\cite{Brodsky:2011ta}
\bibitem{Brodsky:2011ta}
  S.~J.~Brodsky and X.~G.~Wu,
  %Scale Setting Using the Extended Renormalization Group and the Principle of Maximum Conformality: the QCD Coupling Constant at Four Loops,
  Phys.\ Rev.\ D {\bf 85}, 034038 (2012)
  [Phys.\ Rev.\ D {\bf 86}, 079903 (2012)].

%\cite{Brodsky:2011ig}
\bibitem{Brodsky:2011ig}
  S.~J.~Brodsky and L.~Di Giustino,
  %Setting the Renormalization Scale in QCD: The Principle of Maximum Conformality,
  Phys.\ Rev.\ D {\bf 86}, 085026 (2012).

%\cite{Mojaza:2012mf}
\bibitem{Mojaza:2012mf}
  M.~Mojaza, S.~J.~Brodsky and X.~G.~Wu,
  %Systematic All-Orders Method to Eliminate Renormalization-Scale and Scheme Ambiguities in Perturbative QCD,
  Phys.\ Rev.\ Lett.\  {\bf 110}, 192001 (2013).

%\cite{Brodsky:2013vpa}
\bibitem{Brodsky:2013vpa}
  S.~J.~Brodsky, M.~Mojaza and X.~G.~Wu,
  %Systematic Scale-Setting to All Orders: The Principle of Maximum Conformality and Commensurate Scale Relations,
  Phys.\ Rev.\ D {\bf 89}, 014027 (2014).

  %\cite{Brodsky:1982gc}
\bibitem{Brodsky:1982gc}
  S.~J.~Brodsky, G.~P.~Lepage and P.~B.~Mackenzie,
  %On the Elimination of Scale Ambiguities in Perturbative Quantum Chromodynamics,
  Phys.\ Rev.\ D {\bf 28}, 228 (1983).

\bibitem{GellMann:1954fq}
  M.~Gell-Mann and F.~E.~Low,
  %Quantum electrodynamics at small distances,
  Phys.\ Rev.\  {\bf 95}, 1300 (1954).

%\cite{Brodsky:2012ms}
\bibitem{Brodsky:2012ms}
  S.~J.~Brodsky and X.~G.~Wu,
  %Self-Consistency Requirements of the Renormalization Group for Setting the Renormalization Scale,
  Phys.\ Rev.\ D {\bf 86}, 054018 (2012).
  %[arXiv:1208.0700 [hep-ph]].

%\cite{Wu:2014iba}
\bibitem{Wu:2014iba}
  X.~G.~Wu, Y.~Ma, S.~Q.~Wang, H.~B.~Fu, H.~H.~Ma, S.~J.~Brodsky and M.~Mojaza,
  %Renormalization Group Invariance and Optimal QCD Renormalization Scale-Setting,
  Rept.\ Prog.\ Phys.\  {\bf 78}, 126201 (2015).
  %[arXiv:1405.3196 [hep-ph]].

%\cite{Brodsky:1997jk}
\bibitem{Brodsky:1997jk}
  S.~J.~Brodsky and P.~Huet,
  %Aspects of SU(N(c)) gauge theories in the limit of small number of colors,
  Phys.\ Lett.\ B {\bf 417}, 145 (1998).

\bibitem{Wu:2015rga}
  X.~G.~Wu, S.~Q.~Wang and S.~J.~Brodsky,
  %Importance of proper renormalization scale-setting for QCD testing at colliders,
  Front.\ Phys.\ {\bf 11}, 111201 (2016).

%\cite{Brodsky:2012rj}
\bibitem{Brodsky:2012rj}
  S.~J.~Brodsky and X.~G.~Wu,
  %Eliminating the Renormalization Scale Ambiguity for Top-Pair Production Using the Principle of Maximum Conformality,
  Phys.\ Rev.\ Lett.\  {\bf 109}, 042002 (2012).

%\cite{Brodsky:2012sz}
\bibitem{Brodsky:2012sz}
  S.~J.~Brodsky and X.~G.~Wu,
  %Application of the Principle of Maximum Conformality to Top-Pair Production,
  Phys.\ Rev.\ D {\bf 86}, 014021 (2012).

%\cite{Brodsky:2012ik}
\bibitem{Brodsky:2012ik}
  S.~J.~Brodsky and X.~G.~Wu,
  %Application of the Principle of Maximum Conformality to the Top-Quark Forward-Backward Asymmetry at the Tevatron,
  Phys.\ Rev.\ D {\bf 85}, 114040 (2012).

%\cite{Wang:2014sua}
\bibitem{Wang:2014sua}
  S.~Q.~Wang, X.~G.~Wu, Z.~G.~Si and S.~J.~Brodsky,
  %Application of the Principle of Maximum Conformality to the Top-Quark Charge Asymmetry at the LHC,
  Phys.\ Rev.\ D {\bf 90}, 114034 (2014).

%\cite{Wang:2015lna}
\bibitem{Wang:2015lna}
  S.~Q.~Wang, X.~G.~Wu, Z.~G.~Si and S.~J.~Brodsky,
  %Predictions for the Top-Quark Forward-Backward Asymmetry at High Invariant Pair Mass Using the Principle of Maximum Conformality,
  Phys.\ Rev.\ D {\bf 93}, 014004 (2016).
  %[arXiv:1508.03739 [hep-ph]].

%\cite{Nason:1987xz}
\bibitem{Nason:1987xz}
  P.~Nason, S.~Dawson and R.~K.~Ellis,
  %``The Total Cross-Section for the Production of Heavy Quarks in Hadronic Collisions,''
  Nucl.\ Phys.\ B {\bf 303}, 607 (1988).
  %1541 citations counted in INSPIRE as of 01 Dec 2017

%\cite{Nason:1989zy}
\bibitem{Nason:1989zy}
  P.~Nason, S.~Dawson and R.~K.~Ellis,
  %``The One Particle Inclusive Differential Cross-Section for Heavy Quark Production in Hadronic Collisions,''
  Nucl.\ Phys.\ B {\bf 327}, 49 (1989).

%\cite{Beenakker:1988bq}
\bibitem{Beenakker:1988bq}
  W.~Beenakker, H.~Kuijf, W.~L.~van Neerven and J.~Smith,
  %``QCD Corrections to Heavy Quark Production in p anti-p Collisions,''
  Phys.\ Rev.\ D {\bf 40}, 54 (1989).

\bibitem{Beenakker:1990maa}
  W.~Beenakker, W.~L.~van Neerven, R.~Meng, G.~A.~Schuler and J.~Smith,
  %``QCD corrections to heavy quark production in hadron hadron collisions,''
  Nucl.\ Phys.\ B {\bf 351}, 507 (1991).

%\cite{Czakon:2008ii}
\bibitem{Czakon:2008ii}
  M.~Czakon and A.~Mitov,
  %``Inclusive Heavy Flavor Hadroproduction in NLO QCD: The Exact Analytic Result,''
  Nucl.\ Phys.\ B {\bf 824}, 111 (2010).

%\cite{Moch:2008qy}
\bibitem{Moch:2008qy}
  S.~Moch and P.~Uwer,
  %``Theoretical status and prospects for top-quark pair production at hadron colliders,''
  Phys.\ Rev.\ D {\bf 78}, 034003 (2008).

%\cite{Beneke:2011mq}
\bibitem{Beneke:2011mq}
  M.~Beneke, P.~Falgari, S.~Klein and C.~Schwinn,
  %``Hadronic top-quark pair production with NNLL threshold resummation,''
  Nucl.\ Phys.\ B {\bf 855}, 695 (2012).

%\cite{Kidonakis:2010dk}
\bibitem{Kidonakis:2010dk}
  N.~Kidonakis,
  %``Next-to-next-to-leading soft-gluon corrections for the top quark cross section and transverse momentum distribution,''
  Phys.\ Rev.\ D {\bf 82}, 114030 (2010).

%\cite{Baernreuther:2012ws}
\bibitem{Baernreuther:2012ws}
  P.~B?rnreuther, M.~Czakon and A.~Mitov,
  %``Percent Level Precision Physics at the Tevatron: First Genuine NNLO QCD Corrections to $q \bar{q} \to t \bar{t} + X$,''
  Phys.\ Rev.\ Lett.\  {\bf 109}, 132001 (2012).

\bibitem{Czakon:2012pz}
  M.~Czakon and A.~Mitov,
  JHEP {\bf 1301}, 080 (2013).
  %%CITATION = doi:10.1007/JHEP01(2013)080;%%

\bibitem{Czakon:2013goa}
  M.~Czakon, P.~Fiedler and A.~Mitov,
  %Total Top-Quark Pair-Production Cross Section at Hadron Colliders Through $O(¦Á\frac{4}{S})$,
  Phys.\ Rev.\ Lett.\  {\bf 110}, 252004 (2013).

%\cite{Aliev:2010zk}
\bibitem{Aliev:2010zk}
  M.~Aliev, H.~Lacker, U.~Langenfeld, S.~Moch, P.~Uwer and M.~Wiedermann,
  %HATHOR: HAdronic Top and Heavy quarks crOss section calculatoR,
  Comput.\ Phys.\ Commun.\  {\bf 182}, 1034 (2011).
  %[arXiv:1007.1327 [hep-ph]].

%\cite{Czakon:2011xx}
\bibitem{Czakon:2011xx}
  M.~Czakon and A.~Mitov,
  %Top++: A Program for the Calculation of the Top-Pair Cross-Section at Hadron Colliders,
  Comput.\ Phys.\ Commun.\  {\bf 185}, 2930 (2014).
  %[arXiv:1112.5675 [hep-ph]].

%\cite{toppole}
\bibitem{toppole}
  The ATLAS and CMS Collaborations,
  %Combination of ATLAS and CMS results on the mass of the top quark using up to 4.9 fb$^{-1}$ of data,
  ATLAS-CONF-2012-095, CMS-PAS-TOP-12-001.

%\cite{Dulat:2015mca}
\bibitem{Dulat:2015mca}
  S.~Dulat, {\it et al.},
  %New parton distribution functions from a global analysis of quantum chromodynamics,
  Phys.\ Rev.\ D {\bf 93}, 033006 (2016).
  %[arXiv:1506.07443 [hep-ph]].

%\cite{Brodsky:2014yha}
\bibitem{Brodsky:2014yha}
  S.~J.~Brodsky, G.~F.~de Teramond, H.~G.~Dosch and J.~Erlich,
  %Light-Front Holographic QCD and Emerging Confinement,
  Phys.\ Rept.\  {\bf 584}, 1 (2015).
  %[arXiv:1407.8131 [hep-ph]].

%\cite{Czakon:2016dgf}
\bibitem{Czakon:2016dgf}
  M.~Czakon, D.~Heymes and A.~Mitov,
  %Dynamical scales for multi-TeV top-pair production at the LHC,
  JHEP {\bf 1704}, 071 (2017).
  %[arXiv:1606.03350 [hep-ph]].

%\cite{Wang:2016wgw}
\bibitem{Wang:2016wgw}
  S.~Q.~Wang, X.~G.~Wu, S.~J.~Brodsky and M.~Mojaza,
  %Application of the Principle of Maximum Conformality to the Hadroproduction of the Higgs Boson at the LHC,
  Phys.\ Rev.\ D {\bf 94}, 053003 (2016).
  %[arXiv:1605.02572 [hep-ph]].

%\cite{Aaltonen:2013wca}
\bibitem{Aaltonen:2013wca}
  T.~A.~Aaltonen {\it et al.} [CDF and D0 Collaborations],
  %Combination of measurements of the top-quark pair production cross section from the Tevatron Collider,
  Phys.\ Rev.\ D {\bf 89}, 072001 (2014).
  %[arXiv:1309.7570 [hep-ex]].

%\cite{Chatrchyan:2013ual}
\bibitem{Chatrchyan:2013ual}
  S.~Chatrchyan {\it et al.} [CMS Collaboration],
  %Measurement of the $t\bar{t}$ production cross section in the all-jet final state in pp collisions at $\sqrt{s}$ = 7 TeV,
  JHEP {\bf 1305}, 065 (2013).
  %[arXiv:1302.0508 [hep-ex]].

%\cite{Aad:2012vip}
\bibitem{Aad:2012vip}
  G.~Aad {\it et al.} [ATLAS Collaboration],
  %Measurement of the ttbar production cross section in the tau+jets channel using the ATLAS detector,
  Eur.\ Phys.\ J.\ C {\bf 73}, 2328 (2013).
  %[arXiv:1211.7205 [hep-ex]].

%\cite{Chatrchyan:2013kff}
\bibitem{Chatrchyan:2013kff}
  S.~Chatrchyan {\it et al.} [CMS Collaboration],
  %Measurement of the top-antitop production cross section in the tau+jets channel in pp collisions at sqrt(s) = 7 TeV,
  Eur.\ Phys.\ J.\ C {\bf 73}, 2386 (2013).
  %[arXiv:1301.5755 [hep-ex]].

%\cite{Aad:2015dya}
\bibitem{Aad:2015dya}
  G.~Aad {\it et al.} [ATLAS Collaboration],
  %Measurements of the top quark branching ratios into channels with leptons and quarks with the ATLAS detector,
  Phys.\ Rev.\ D {\bf 92}, 072005 (2015).
  %[arXiv:1506.05074 [hep-ex]].

%\cite{Chatrchyan:2012vs}
\bibitem{Chatrchyan:2012vs}
  S.~Chatrchyan {\it et al.} [CMS Collaboration],
  %Measurement of the top quark pair production cross section in $pp$ collisions at $\sqrt{s} = 7$ TeV in dilepton final states containing a $\tau$,
  Phys.\ Rev.\ D {\bf 85}, 112007 (2012).
  %[arXiv:1203.6810 [hep-ex]].

%\cite{Aad:2012qf}
\bibitem{Aad:2012qf}
  G.~Aad {\it et al.} [ATLAS Collaboration],
  %Measurement of the top quark pair production cross-section with ATLAS in the single lepton channel,
  Phys.\ Lett.\ B {\bf 711}, 244 (2012).
  %[arXiv:1201.1889 [hep-ex]].

%\cite{Chatrchyan:2016abc}
\bibitem{Chatrchyan:2016abc}
  S.~Chatrchyan {\it et al.} [CMS Collaboration],
  %Measurements of the t-tbar production cross section in lepton+jets final states in pp collisions at 8 TeV and ratio of 8 to 7 TeV cross sections,
  Eur.\ Phys. \ J. \ C {\bf 77}, 15 (2017).
%  [arXiv:1602.09024 [hep-ex]].

%\cite{Khachatryan:2015fwh}
\bibitem{Khachatryan:2015fwh}
  V.~Khachatryan {\it et al.} [CMS Collaboration],
  %Measurement of the $\mathrm{t}\overline{{\mathrm{t}}}$ production cross section in the all-jets final state in pp collisions at $\sqrt{s}=8$ $\,\text {TeV}$,
  Eur.\ Phys.\ J.\ C {\bf 76}, 128 (2016).
  %[arXiv:1509.06076 [hep-ex]].

%\cite{Khachatryan:2014loa}
\bibitem{Khachatryan:2014loa}
  V.~Khachatryan {\it et al.} [CMS Collaboration],
  %Measurement of the $t \bar t$ production cross section in $pp$ collisions at $\sqrt s = 8$ TeV in dilepton final states containing one $\tau$ lepton,
  Phys.\ Lett.\ B {\bf 739}, 23 (2014).
  %[arXiv:1407.6643 [hep-ex]].

%\cite{Aad:2015pga}
\bibitem{Aad:2015pga}
  G.~Aad {\it et al.} [ATLAS Collaboration],
  %Measurement of the top pair production cross section in 8 TeV proton-proton collisions using kinematic information in the lepton+jets final state with ATLAS,
  Phys.\ Rev.\ D {\bf 91}, 112013 (2015).
  %[arXiv:1504.04251 [hep-ex]].

%\cite{Khachatryan:2016kzg}
\bibitem{Khachatryan:2016kzg}
  V.~Khachatryan {\it et al.} [CMS Collaboration],
  %Measurement of the $\mathrm{t \bar{t}}$ production cross section using events in the $\mathrm{e} \mu$ final state in pp collisions at $\sqrt{s} = $ 13 TeV,
  arXiv:1611.04040 [hep-ex].

%\cite{Khachatryan:2015uqb}
\bibitem{Khachatryan:2015uqb}
  V.~Khachatryan {\it et al.} [CMS Collaboration],
  %Measurement of the top quark pair production cross section in proton-proton collisions at $\sqrt(s) =$ 13 TeV,
  Phys.\ Rev.\ Lett.\  {\bf 116}, 052002 (2016).
  %[arXiv:1510.05302 [hep-ex]].

%\cite{Aaboud:2015AAAA}
\bibitem{Aaboud:2015AAAA}
  M.~Aaboud {\it et al.} [ATLAS Collaboration],
  %Measurements of the $t\bar{t}$ production cross-section in the dilepton and lepton-plus-jets channels and of the ratio of the $t\bar{t}$ and Z boson cross-sections in pp collisions at $\sqrt{S}=13$ TeV with the ATLAS detector,
  ATLAS-CONF-2015-049.

%\cite{Aaboud:2016pbd}
\bibitem{Aaboud:2016pbd}
  M.~Aaboud {\it et al.} [ATLAS Collaboration],
  %Measurement of the $t\bar{t}$ production cross-section using $e\mu$ events with b-tagged jets in pp collisions at $\sqrt{s}$=13 TeV with the ATLAS detector,
  Phys.\ Lett.\ B {\bf 761}, 136 (2016).
  %[arXiv:1606.02699 [hep-ex]].

%\cite{Abazov:2009si}
\bibitem{Abazov:2009si}
  V.~M.~Abazov {\it et al.} [D0 Collaboration],
  %Measurement of the t anti-t production cross section and top quark mass extraction using dilepton events in p anti-p collisions,
  Phys.\ Lett.\ B {\bf 679}, 177 (2009).
  %[arXiv:0901.2137 [hep-ex]].

%\cite{Abazov:2011cq}
\bibitem{Abazov:2011cq}
  V.~M.~Abazov {\it et al.} [D0 Collaboration],
  %Measurement of the $t\bar{t}$ production cross section using dilepton events in $p\bar{p}$ collisions,
  Phys.\ Lett.\ B {\bf 704}, 403 (2011).
  %[arXiv:1105.5384 [hep-ex]].

%\cite{Abazov:2011mi}
\bibitem{Abazov:2011mi}
  V.~M.~Abazov {\it et al.} [D0 Collaboration],
  %Measurement of the top quark pair production cross section in the lepton+jets channel in proton-antiproton collisions at $\sqrt{s}$=1.96 TeV,
  Phys.\ Rev.\ D {\bf 84}, 012008 (2011).
  %[arXiv:1101.0124 [hep-ex]].

%\cite{Abazov:2011pta}
\bibitem{Abazov:2011pta}
  V.~M.~Abazov {\it et al.} [D0 Collaboration],
  %Determination of the pole and $\overline{MS}$ masses of the top quark from the $t\bar{t}$ cross section,
  Phys.\ Lett.\ B {\bf 703}, 422 (2011).
  %[arXiv:1104.2887 [hep-ex]].

%\cite{Beneke:2012wb}
\bibitem{Beneke:2012wb}
  M.~Beneke, P.~Falgari, S.~Klein, J.~Piclum, C.~Schwinn, M.~Ubiali and F.~Yan,
  %Inclusive Top-Pair Production Phenomenology with TOPIXS,
  JHEP {\bf 1207}, 194 (2012).
  %[arXiv:1206.2454 [hep-ph]].

%\cite{Chatrchyan:2012bra}
\bibitem{Chatrchyan:2012bra}
  S.~Chatrchyan {\it et al.} [CMS Collaboration],
  %Measurement of the $t\bar{t}$ production cross section in the dilepton channel in $pp$ collisions at $\sqrt{s}=7$ TeV,
  JHEP {\bf 1211}, 067 (2012).
  %[arXiv:1208.2671 [hep-ex]].

%\cite{Chatrchyan:2012ria}
\bibitem{Chatrchyan:2012ria}
  S.~Chatrchyan {\it et al.} [CMS Collaboration],
  %Measurement of the $t\bar{t}$ production cross section in $pp$ collisions at $\sqrt{s}=7$ TeV with lepton + jets final states,
  Phys.\ Lett.\ B {\bf 720}, 83 (2013).
  %[arXiv:1212.6682 [hep-ex]].

%\cite{Chatrchyan:2013faa}
\bibitem{Chatrchyan:2013faa}
  S.~Chatrchyan {\it et al.} [CMS Collaboration],
  %Measurement of the $t \bar{t}$ production cross section in the dilepton channel in pp collisions at $\sqrt{s}$ = 8 TeV,
  JHEP {\bf 1402}, 024 (2014)
  Erratum: [JHEP {\bf 1402}, 102 (2014)].
  %[arXiv:1312.7582 [hep-ex], arXiv:1312.7582].

%\cite{Marquard:2015qpa}
\bibitem{Marquard:2015qpa}
  P.~Marquard, A.~V.~Smirnov, V.~A.~Smirnov and M.~Steinhauser,
  %Quark Mass Relations to Four-Loop Order in Perturbative QCD,
  Phys.\ Rev.\ Lett.\  {\bf 114}, 142002 (2015).
  %[arXiv:1502.01030 [hep-ph]].

\bibitem{Kataev:2015gvt}
  A.~L.~Kataev and V.~S.~Molokoedov,
  %On the flavour dependence of the $\mathcal{O}(\alpha_s^4)$ correction to the relation between running and pole heavy quark masses,
  Eur.\ Phys.\ J.\ Plus {\bf 131}, 271 (2016).

%\cite{Abazov:2009ae}
\bibitem{Abazov:2009ae}
  V.~M.~Abazov {\it et al.} [D0 Collaboration],
  %Combination of t anti-t cross section measurements and constraints on the mass of the top quark and its decays into charged Higgs bosons,
  Phys.\ Rev.\ D {\bf 80}, 071102 (2009).
  %[arXiv:0903.5525 [hep-ex]].

%\cite{Aad:2015waa}
\bibitem{Aad:2015waa}
  G.~Aad {\it et al.} [ATLAS Collaboration],
  %Determination of the top-quark pole mass using $ t\overline{t} $ + 1-jet events collected with the ATLAS experiment in 7 TeV pp collisions,
  JHEP {\bf 1510}, 121 (2015).
  %[arXiv:1507.01769 [hep-ex]].

%\cite{Baak:2012kk}
\bibitem{Baak:2012kk}
  M.~Baak {\it et al.},
  %The Electroweak Fit of the Standard Model after the Discovery of a New Boson at the LHC,
  Eur.\ Phys.\ J.\ C {\bf 72}, 2205 (2012).
  %[arXiv:1209.2716 [hep-ph]].

%\cite{Khachatryan:2016pek}
\bibitem{Khachatryan:2016pek}
  V.~Khachatryan {\it et al.} [CMS Collaboration],
  %`Measurement of the mass of the top quark in decays with a $J/\psi$ meson in pp collisions at 8 TeV,
  JHEP {\bf 1612}, 123 (2016).
  %doi:10.1007/JHEP12(2016)123
  %[arXiv:1608.03560 [hep-ex]].

%\cite{2017mlepton}
\bibitem{2017mlepton}
  G.~Aad {\it et al.} [ATLAS Collaboration],
  %Measurement of lepton differential distributions and the top quark mass in $t\bar{t}$ production in pp collisions at $\sqrt{S}$ = 8 TeV with the ATLAS detector,
  ATLAS-CONF-2017-044.

%\cite{Aaboud:2016igd}
\bibitem{Aaboud:2016igd}
  M.~Aaboud {\it et al.} [ATLAS Collaboration],
  %Measurement of the top quark mass in the $t\bar{t}\to$ dilepton channel from $\sqrt{s}=8$ TeV ATLAS data,
  Phys.\ Lett.\ B {\bf 761}, 350 (2016).
  %doi:10.1016/j.physletb.2016.08.042
  %[arXiv:1606.02179 [hep-ex]].

%\cite{Khachatryan:2015hba}
\bibitem{Khachatryan:2015hba}
  V.~Khachatryan {\it et al.} [CMS Collaboration],
  %Measurement of the top quark mass using proton-proton data at ${\sqrt{(s)}}$ = 7 and 8 TeV,
  Phys.\ Rev.\ D {\bf 93}, 072004 (2016).
  %doi:10.1103/PhysRevD.93.072004
  %[arXiv:1509.04044 [hep-ex]].


\end{thebibliography}
\end{document}